\documentclass[prl,aps,twocolumn,preprintnumbers,floats,amsmath,amssymb]{revtex4}

\usepackage{graphicx}
\usepackage{dcolumn}
\usepackage{bm}
\usepackage[latin1]{inputenc}

\def \be{\begin{equation}}
\def \ee{\end{equation}}

\def\moy#1{\left\langle #1 \right\rangle}

\begin{document}

\title{Hybrid phase at the quantum melting of the Wigner crystal}

\author{Houman Falakshahi and Xavier Waintal}
\affiliation{Nanoelectronic Group, Service de Physique de l'Etat Condensé,
CEA Saclay F-91191 Gif-sur-Yvette Cedex, France\\}


\begin{abstract}
We study the quantum melting of the two-dimensional Wigner crystal
using a fixed node quantum Monte-Carlo approach. In addition to the
two already known phases (Fermi liquid at large density and
Wigner crystal at low density), we find a third stable phase at intermediate 
values of the density. The third phase has hybrid behaviors in between a 
liquid and a solid. This hybrid phase has the nodal structure 
of a Slater determinant constructed out of the bands of a
triangular lattice. 
\end{abstract}
\maketitle
The physics of a system of $N$ electrons confined on a two dimensional surface $S$
is a textbook problem at the root of a very large body of literature.
Two competing energies, electrostatic and kinetic, give rise to a rich phase
diagram. The physics is controlled by the dimensionless parameter 
$r_s=m^* e^2/(\hbar^2\epsilon \sqrt{\pi n})$ which is the ratio
of the average distance between electrons over the effective Bohr radius
($e$ is the electronic charge, $\epsilon$ the dielectric constant, $m^*$ the
effective mass and $n=N/S$ the electronic density).
At large density (low $r_s$), the kinetic energy dominates and the system is in a 
Fermi liquid phase~\cite{pines}. Since the work of Wigner~\cite{wigner} in 1934, 
it is also known that at low density (large $r_s$), the Coulomb repulsion dominates and
the electrons crystallize onto a (Wigner) triangular crystal~\cite{wigner,maradudin}.
 In their pioneering work in 1989, Tanatar and Ceperley~\cite{tanatar_ceperley} 
were able to locate that the quantum melting of the crystal occurs for a critical value
of $r_s\approx 37\pm 5$. There work, which used a Fixed Node Quantum Monte-Carlo~\cite{alder} 
(FN-QMC)technique, was followed by more precise numerics~\cite{senatore} and a better
description of the liquid phase~\cite{attaccalite,kwon} that included backflow corrections.

This simple picture of a, presumably first order, direct transition between the
solid and the liquid phase 
is to be contrasted with other aspects of the physics of the Wigner crystal 
which show more complex behaviors. For instance, its magnetism 
is believed to include a spin liquid phase in addition to the ferromagnetic phase found 
at very large $r_s$ \cite{bernu_ceperley}. The fermionic statistics of the electrons
is also known to play a crucial role for $r_s\le 60$ where the melting of the 
bosonic Wigner crystal occurs~\cite{bernu2}. Also, the classical melting 
\cite{williams,kosterlitz,halperin,platzman} (as a function of temperature)
occurs in two steps.
The  system first looses its translational order but retains some orientational order (hexatic
phase~\cite{halperin}) while at higher temperature, all order disappears.
The possibility that the melting of a quantum crystal would also take place
in two steps, leading to a highly correlated intermediate phase has been discussed
as early as 1969 by Andreev and Lifshitz~\cite{andreev-lifshitz}, 
who proposed that a liquid of defects would exist together with the crystal state.
This proposal has been revisited recently in small systems using exact diagonalization
techniques~\cite{katomeris,nemeth} as well as in screened systems~\cite{spivak}. 
\begin{figure}
\includegraphics[width=8cm]{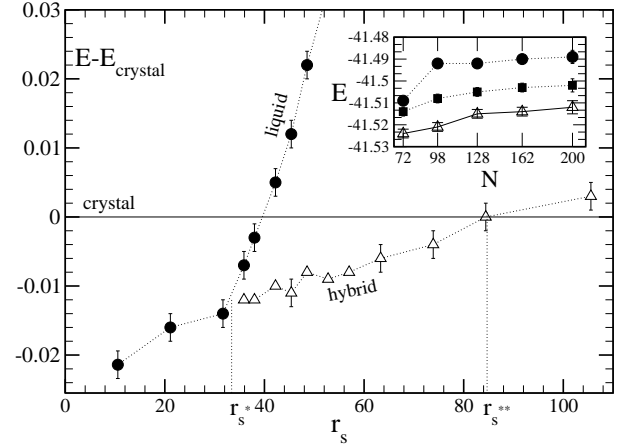}
\caption{\label{stability} Energy difference  $E_{\rm liquid} -E_{\rm crystal}$ (circles) and
$E_{\rm hybrid} -E_{\rm crystal}$ (triangles) as a function of $r_s$ for $72$ electrons 
in a $48\times 84$ grid. Inset: Energies of the three phases at $\nu=1/56$ and
$U=20$ ($r_s\approx 42.2$) as a function of the number of particles up to $200$ electrons in a $80\times 140$ grid.}
\end{figure}

In this letter, we study a new phase which is a hybrid of a liquid and a solid
using a FN-QMC technique similar to the one used in~\cite{tanatar_ceperley}.
The FN-QMC approach is a very powerful tool to tackle this problem, but it is of primary
importance to understand the nature of the approximations which it involves. The
method lies half way between a black box and a variational approach. Technically, 
the FN-QMC algorithm is fed with a wave-function called the {\it guiding wave function} 
(GWF) that has to be given explicitly, and that should be close to the 
ground state of the system. The FN-QMC algorithm modifies the GWF to become 
as close as possible to the ground state of the system,
given the constraint that {\it the sign of the wave function
remains unchanged} at every point of the Hilbert space. The method
gives the best wave function for a given structure of the nodes 
of the GWF and is in this sense variational~\cite{haaf}.
Our main result is summarized in the
stability diagram Fig.~\ref{stability} where the energies of the different phases 
(i.e. associated with the different GWFs) are plotted 
as a function of $r_s$. The hybrid phase is found to be stable in the (critical) region $r_s^*<r_s<r_s^{**}$ 
with $r_s^*\approx 30$ and $r_s^{**}\approx 80$. 

{\it Model.} We consider a system of $N$ spinless electrons on a square $L_x\times L_y$ grid 
with periodic boundary conditions whose Hamiltonian is given by,
\be
\label{eq:model}
H=-t\sum_{\langle\vec r,\vec r'\rangle}c_{\vec r}^\dagger c_{\vec r'}
+\frac{U}{2} \sum_{\vec r\ne\vec r'} V(\vec r-\vec r') n_{\vec r} n_{\vec r'}  + \lambda.
\ee
The operator $c_{\vec r}^\dagger$ ($c_{\vec r}$) creates (destroys) an electron on
point $\vec r$ with the standard anticommutation relation rules. 
The sum $\sum_{\langle\vec r,\vec r'\rangle}$ is done on the nearest
neighbor points on the grid and $t$ is the corresponding hopping amplitude. 
The density operator reads $n_{\vec r}=c_{\vec r}^\dagger c_{\vec r}$.  
$U$ is the effective strength of the interaction. 
The two body interaction $V(\vec r)$ is obtained from the bare Coulomb interaction using the
Ewald summation techniques to avoid finite size effects, and reads
\be
V(\vec r) = \sum_{\vec L} \frac{1}{|\vec r +\vec L|} {\rm Erfc}(k_c |\vec r +\vec L|)
\ee
$$
+\frac{2\pi}{L_x L_y} \sum_{\vec K\ne \vec 0} 
\frac{1}{|\vec K|} 
{\rm Erfc}(|\vec K|/(2 k_c))\cos (\vec K\cdot\vec r) .
$$
In the previous equation, $k_c$ is a (irrelevant) cut off. The vector $\vec L$
takes discrete values $\vec L= (n_x L_x,n_y L_y)$ with $n_x$ and $n_y$ integer numbers.
The vector $\vec K$ also takes discrete values, $\vec K= (\frac{2 \pi}{L_x} n_x,
\frac{2 \pi}{L_y} n_y)$ and $(n_x,n_y)\ne (0,0)$. The complementary error function is
${\rm Erfc}(r)=\frac{2}{\sqrt{\pi}} \int_x^\infty e^{-t^2} dt$.
In order to assure electrostatic neutrality we add a positive continuous background 
(the positive background charges are not put on the grid but lie in the continuum). 
The constant term $\lambda$ hence reads,
\be
\lambda/N = 4 t + U V(\vec 0) -U \nu \frac{2\sqrt{\pi}}{k_c} - U \frac{2 k_c}{\sqrt{\pi}}
\ee
where $\nu=\frac{N}{L_x L_y}$ is the average electronic density. All the energies in the
problem are measured in unit of $N 2\pi \nu t$. The $r_s$ parameter for this model reads,
$r_s=U/(2 t \sqrt{\pi \nu})$.
When $\nu \ll 1$ the role of the grid becomes irrelevant and Eq.(\ref{eq:model})
tends toward the continuous model studied in Ref~\cite{tanatar_ceperley} provided 
our energies are multiplied by $2/r_s^2$. As we shall see however, the presence of the
grid gives the possibility of constructing new types of GWF. In our numerics we have 
used $\nu=1/56$ and $\nu=1/780$. Standard two dimensional gas in GaAs heterostructures 
where the underlying grid is given by the Ga and As atoms correspond to 
$\nu\approx 1/1000$ or $\nu\approx 1/10000$ for the most diluted ones. 
In order for the Wigner crystal to fit into the system without 
distortion, we chose $L_y\approx\sqrt{3}L_x$ and $N=2P^2$ with $P$ integer.
The electrons in our study are fully spin polarized which corresponds to a system 
with a strong in plane magnetic field. However our results also extend to zero field systems since at $r_s\ge 20$ the polarized fluid is more stable than the non polarized 
one~\cite{senatore,attaccalite}. Last, we have added a very small (irrelevant)
disorder to the system in order to lift the degeneracies of the non interacting problem.

{\it FN-QMC Method.} The operator $e^{-H t}$ is 
applied stochastically to an initial GWF in order to project it to the
exact ground state. Our implementation is based on the Green Function Monte Carlo 
for lattice Hamiltonians introduced in~\cite{trivedi}. Important sampling~\cite{kalos} 
and Fixed Node are implemented as in \cite{haaf} by
replacing $H$ by an effective Hamiltonian $H_{\rm FN}$ that depends
on the GWF. $H_{\rm FN}$ forbids the sign of the 
wave-function to change. The energies calculated with $H_{\rm FN}$ are larger than the one
of the true ground state but smaller than the variational energy associated with the guiding
wave-function~\cite{haaf}. 
At $\nu\ll 1$ the technique is equivalent to the continuous fixed node 
diffusive Monte-Carlo used in~\cite{tanatar_ceperley}.
The algorithm to update the Slater determinants can be found in~\cite{chester}. 
By sampling directly the time spent by the walkers at one point of the Hilbert space using
the algorithm described in \cite{trivedi} we can use arbitrary small time steps
and effectively work in continuous (imaginary) time. Instead of using branching, 
the control of the walkers population is done
using a fixed number of walkers and the reconfiguration algorithm introduced by 
Sorella~\cite{sorella}. This algorithm allows to avoid the bias introduced 
in the branching technique by 
artificially controlling the walker population. Quantum averages of physical quantities 
$\langle\dots\rangle$ are calculated using the
forward walking technique~\cite{sorella}, and hence do not suffer from the bias of mixed
estimates. A typical point for $72$ particles involves $20$ independent Monte-Carlo runs
with $5000$ walkers each.

{\it Guiding wave functions.} The GWFs used in our calculations 
are Slater determinants multiplied by Jastrow functions,
\be
\Psi(\vec r_1,\vec r_2 ...\vec r_N)=
{\rm Det\ }[\phi_i(\vec r_j)]\times \prod_{i<j} J(| \vec r_i - \vec r_j|). 
\ee
The Jastrow part takes Coulomb interaction into account  
by introducing correlations between electrons. It 
has no nodes, and thus is irrelevant in the FN-QMC results.
We use modified Yukawa functions\cite{stevens},
$J(r)=\frac{A(r_s)}{r}(1-e^{-B(r_s) r})$, where the distances are
measured in unit of the average distance between nearest particles
and $A(r_s)$ and $B(r_s)$ are (optimized) variational parameters. We checked that the
FN-QMC results are not sensitive to the choice of the Jastrow function.
The Slater determinant of one-body wave functions, ${\rm Det\ }[\phi_i(\vec r_j)]$
enforces the antisymmetric nature of the fermionic wave function and is 
responsible for the nodal structure of the GWF. The GWF used in the literature
are constructed out of plane waves $\phi_i(\vec r_j)\propto e^{i\vec k_i \cdot \vec r_j}$
for the liquid GWF $\Psi_{\rm liq}$ and localized orbitals
$\phi_i(\vec r_j)\propto e^{-(\vec r_j-\vec u_i)^2/d_0^2}$ for the crystal GWF $\Psi_{\rm cry}$.
Here the $\vec u_i$ with $i\in\{1\dots N\}$ stand for the positions of the electrons
in the classical crystal and $d_0$ is a variational parameter. 
$\Psi_{\rm liq}$ ($\Psi_{\rm cry}$) provides the exact ground state of $H$ at very
large (low) density. 

{\it Hybrid GWF.} Below we give the detailed construction of a new GWF, $\Psi_{\rm hyb}$, 
such that the $\phi_i(\vec r_j)$ are the Bloch states of a triangular crystal.
First, an effective one-body Hamiltonian $H_{\rm eff}$ is constructed for an effective hole
in a periodic potential given by a classical Wigner crystal, 
\be
H_{\rm eff}=-t\sum_{\langle\vec r,\vec r'\rangle}c_{\vec r}^\dagger c_{\vec r'}
-U^* \sum_{\vec r} W(\vec r) n_{\vec r} 
\ee
where the one-body potential is $W(\vec r)=\sum_{i=1}^N V(\vec r -\vec u_i)$.
The singularity of $W(\vec r)$ at $\vec r=\vec u_i$ has been removed by setting
$W(\vec u_i)\equiv W(\vec u_i +(1,0))$ and we checked that our results are unaffected
by this choice. In a second step, we take advantage of the presence of the underlying grid
and $H_{\rm eff}$ is numerically diagonalized using Lanczos algorithm. The $N$ orbitals
of lowest energy $\phi_i(\vec r)$ ($1\le i\le N$) are then used to construct the Slater
determinant. $U^*$ is a variational parameter.

The underlying idea behind the construction of $\Psi_{\rm hyb}$ is to put on the same
level the melting of the Wigner crystal in real space (as the density is increased)
and the destruction of the Fermi sea in momentum space (as the density is decreased).
$\Psi_{\rm hyb}$ allows for an interpolation between momentum space ($U^*=0$) 
and real space ($U^*\gg 1$). However, it never properly describes the Wigner crystal,
since according to Bloch theorem, the $\phi_i(\vec r_j)$ are always delocalized states 
(that can be concentrated around the $\vec u_i$'s but that are delocalized anyway).
The available values of momentum $\vec k$ are taken within the first Brillouin zone, and hence,
the liquid-hybrid transition can be viewed as an instability of the shape of the
Fermi surface that goes from a circular to a hexagonal form. 
The symmetry is broken at this transition, 
but it is only in a second step that larger values of $|\vec k|$ will come into play, 
allowing the $\phi_i(\vec r_j)$ to get localized and the actual crystallization 
to take place. 
This transition in two steps, where first the direction of $\vec k$ and 
secondly its absolute value are affected, is
reminiscent of the hexatic phase predicted in the classical melting.

\begin{figure}
\includegraphics[width=8cm]{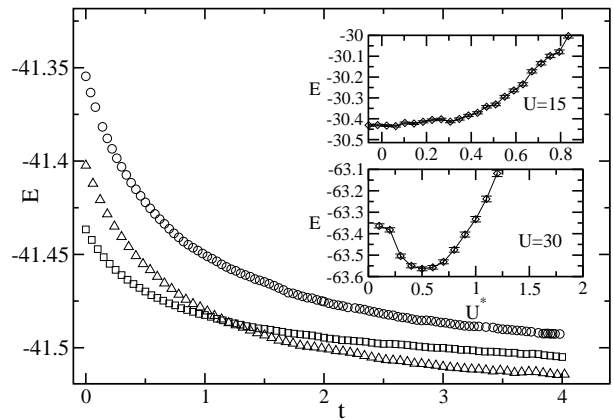}
\caption{\label{technique}
 Energy of the liquid (circle, $A=6.0$, $B=2.25$), 
hybrid (triangle, $A=4.9$,  $B=2.58$,  $U^*=0.3$) and 
crystal (square, $A=4.9$,  $B=2.5$,  $d_0=2.95$) phase as a function of imaginary time $t$
for  $72$ electrons in a $48\times 84$ grid at $U=20$ ($r_s\approx 42.2$). Inset:
variational energy of the hybrid phase as a function of $U^*$ at $U=15$ 
($r_s\approx 31.67$ upper panel, $A=4.7$, $B=2.3$) and $U=30$ ($r_s\approx 63.33$ lower panel, $A=5.7$, $B=2.9$)
}
\end{figure}

\begin{figure}
\includegraphics[width=8cm]{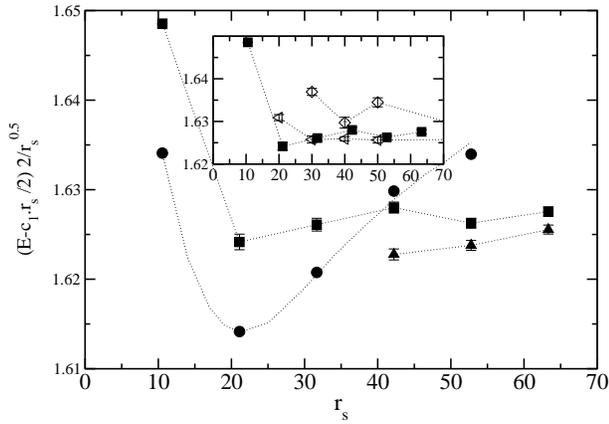}
\caption{\label{comparaison}
$\left[2 E(r_s)- c_1 r_s\right]/r_s^{1/2}$ as a function of $r_s$
for $72$ electrons in a $180 \times 312$ grid. The curves
for the liquid (circles), crystal (squares) and hybrid phase (triangle) can be compared
directly with Fig.2 of \cite{tanatar_ceperley}.
Inset: same thing for the crystal phase only. The curves show our data for $72$ electrons in a $180 \times 312$ 
grid (squares) the data of Ref.~\cite{tanatar_ceperley} for $56$ electrons (diamonds) and the data of Ref.~\cite{senatore} for $56$ electrons (triangles).
}
\end{figure}

{\it Stability of the hybrid phase.}
Fig.\ref{stability}
shows the energy differences $E_{\rm liq} -E_{\rm cry}$ and
$E_{\rm hyb} -E_{\rm cry}$ as a function of $r_s$ for a system of 72 electrons
in a $48\times 84$ grid. These energy differences are very small, less than
0.1\% of the total energy of the sytem. $r_s\approx 40$ where $E_{\rm liq} -E_{\rm cry}\approx 0$
would be the critical value or $r_s$ in the absence of the hybrid phase~\cite{tanatar_ceperley}.
However, we find that for $30<r_s<80$, the hybrid phase has a smaller energy than
both the liquid and the solid phase. Around $r_s^*\approx 30$ we find a jump of $U^*$ from
zero to $U^*=0.3$, see the inset of Fig.~\ref{technique}. 
$U^*/U\approx 0.015$ up to $r_s^{**}\approx 80$, above which
the crystal phase become more stable than the hybrid phase. Finite $N$ corrections 
shown in the inset of Fig.\ref{stability} at $r_s\approx 42.2$ do not perturb the
previous picture. We note that although the variational energy of the hybrid phase
is lower than the one of the liquid it is still higher than the crystal variational
energy. The FN-QMC treatment is necessary to show the stability of the hybrid phase, as
shown in Fig.~\ref{technique}. To make contact with the calculations of \cite{tanatar_ceperley,
senatore}, we have repeated these calculations for a more diluted system $\nu=1/780$ where
the role of the underlying grid is negligible. The results are plotted in Fig.~\ref{comparaison}
in the same way as Fig.2 of \cite{tanatar_ceperley} (with $c_1=-2.2122$). In the inset of
Fig.~\ref{comparaison} we have reported the Wigner crystal data of \cite{tanatar_ceperley} and \cite{senatore} for comparison and find a good quantitative agreement with the latter. 
We note that finite $N$ corrections would raise the energies by an amount $\sim 0.01$
(see the inset of Fig.~\ref{stability}) and are difficult to evaluate, especially in 
the absence of an analytical ansatz for the hybrid phase.  

\begin{figure}
\vglue +0.45cm
\includegraphics[height=6cm,width=4cm]{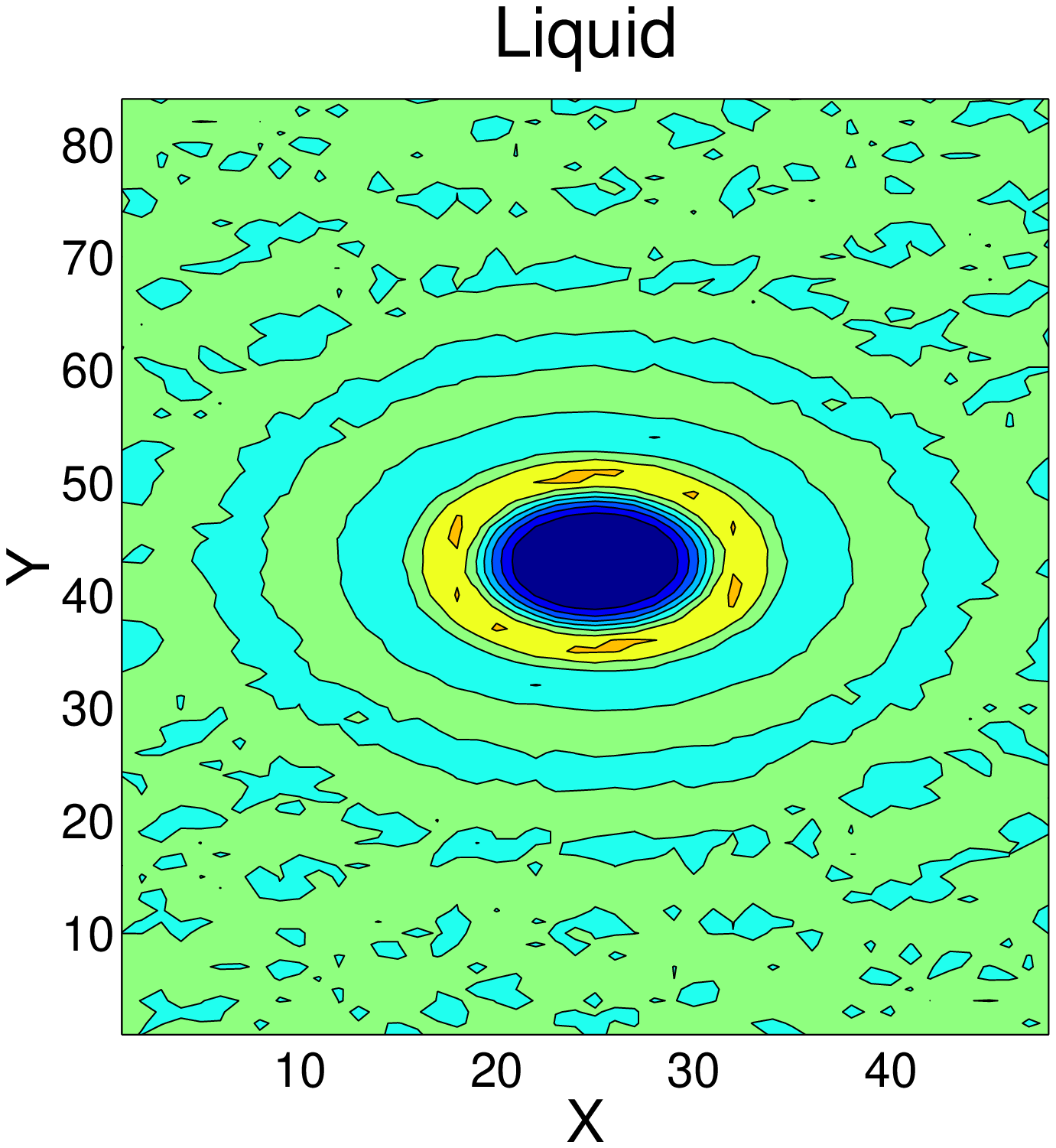}
\includegraphics[height=6cm,width=4cm]{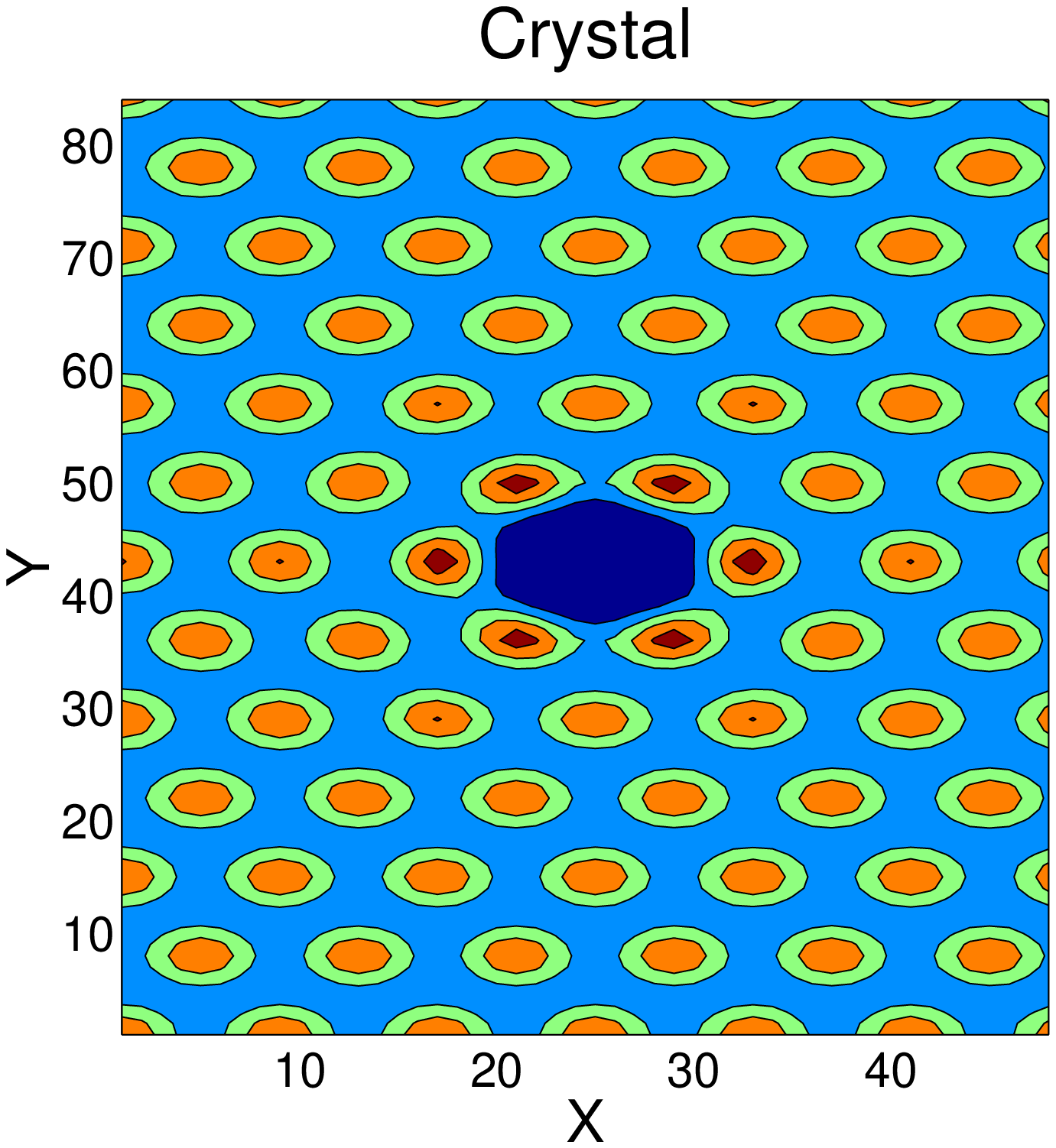}
\includegraphics[height=6cm,width=4.9cm]{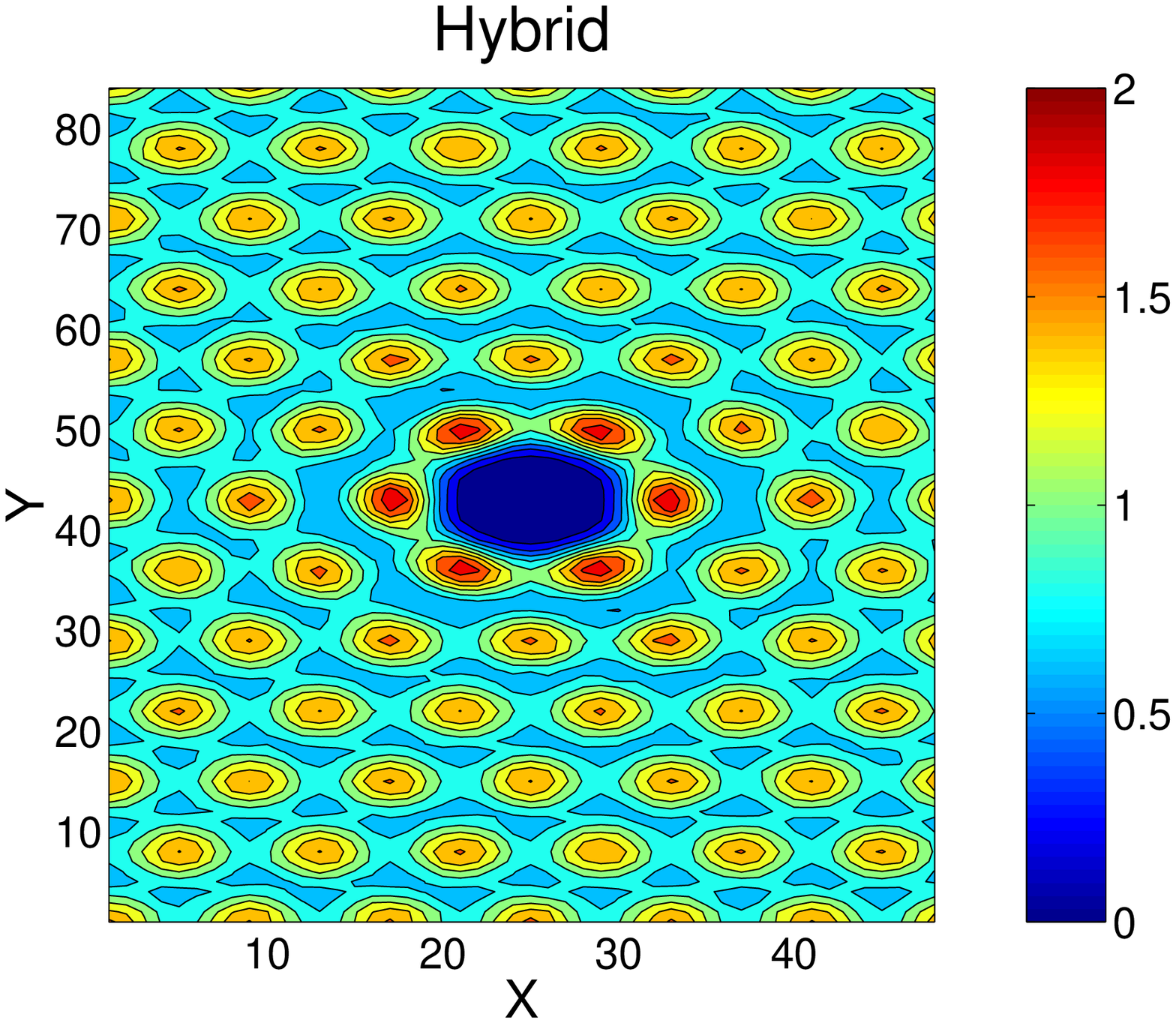}
\caption{\label{g-r} Density-density correlation function at for a system 
of $72$ electrons in a $48\times 84$ grid. $g(\vec r)$ measures the probability of finding a particle 
in $(X,Y)$ knowing that one electron lies in the middle of the sample. From left to right the liquid, crystal and hybrid phases are represented at $U=20$ ($r_s\approx42.2$). }
\end{figure}

{\it Nature of the hybrid phase.} It is important to realize that, as described above,
the nature of the intermediate phase is by construction something hybrid, being
made of (delocalized) Bloch waves, yet having already the symmetry of the Wigner crystal.
More insight can be gained by computing the electronic density $\langle n_{\vec r}\rangle$
(not shown) which is the superposition
of peaks at the classical positions (the $\vec u_i$'s) of the electrons in the crystal 
over a small background. The   background is found to contain 
approximately 35\% of the electrons while the rest lies in the peaks of the crystal.
In that sense, the crystal part of the hybrid phase contains fewer electrons than sites, hence allowing exchange to take place. This is to be contrasted with the conjecture of 
Ref.~\cite{andreev-lifshitz} where a crystal with fewer sites than electrons was predicted.
Although the total energy of the hybrid phase is below those of the liquid and crystal, 
both its kinetic and electrostatic energies lie in between those of the liquid and solid.
Fig.~\ref{g-r} shows the density-density correlation function (roughly measuring
the probability of finding an electron at point $\vec r$ knowing that an electron is at point $\vec 0$), $g(\vec r)=\frac{L_xL_y}{N(N-1)}\moy{c_{\vec r}^\dagger c_{\vec 0}^\dagger c_{\vec 0} c_{\vec r}}$ for the three phases. $g_{\rm hyb}(\vec r)$ for the hybrid phase is intermediate 
between a liquid and a crystal. The value of $g_{\rm hyb}(\vec r)$ at its peaks is only twice 
as big as in the valley to be compared to a factor $15$ at $r_s=100$. 
In fact a very good fit is obtained with 
$g_{\rm hyb}(\vec r) \approx 0.35 \ g_{\rm liq}(\vec r)+0.65 \ g_{\rm cry}(\vec r)$.  

To conclude, we find that a new quantum phase is to be expected instead of a direct melting
of the Wigner crystal. This intermediate phase, whose physical properties remain to be 
investigated in more depths, has hybrid behaviour between those of a solid and a liquid.

{\it Acknowledgement.} We are grateful to D. L'H\^ote, P. Roche, J. Segala and 
F.I.B. Williams for useful discussions. Special thanks to J.-L. Pichard to whom
we are indebted for many stimulating discussions.

\end{document}